\documentclass{webofc}
\usepackage[varg]{txfonts}   
\usepackage{mathtools}
\usepackage{amsmath}
\usepackage{xcolor}
\usepackage{hyperref}
\newcommand{\ud}{\mathrm{d}}
\begin{document}
\title{Shannon entropy of fragmentation functions and their Kullback-Leibler divergence to parton pdfs}
%
%

\author{\firstname{Guillermo} \lastname{Benito Calvi\~no}\inst{1}\fnsep \and
        \firstname{Javier} \lastname{Garc\'{\i}a Olivares}\inst{1}\fnsep\and
        \firstname{Felipe J.} \lastname{Llanes-Estrada}\inst{1}\fnsep\thanks{\email{fllanes@fis.ucm.es}}
}

\institute{F\'{\i}sica Te\'orica \& IPARCOS, Fac. Ciencias F\'{\i}sicas,
           Plaza de las Ciencias 1, 28040 Madrid, Spain.
          }

\abstract{
The flow of information in high-energy collisions has been recently investigated by various groups: this includes the entanglement entropy of the proton becoming classical information entropy of pdfs, jet splitting affecting entropy, or the entropy distribution in hadron decays. Here we examine fragmentation functions in this context, including their entropy as probability distributions, and propose it as one convenient number to characterize progress in their extraction. We also use the Kullback-Leibler divergence to examine relations between FFs and pdfs such as that of Barone, Drago and Ma.
}
\maketitle
\section{Introduction} \label{intro}

\begin{figure}[h]
	\centering
		\includegraphics[width=0.7\columnwidth]{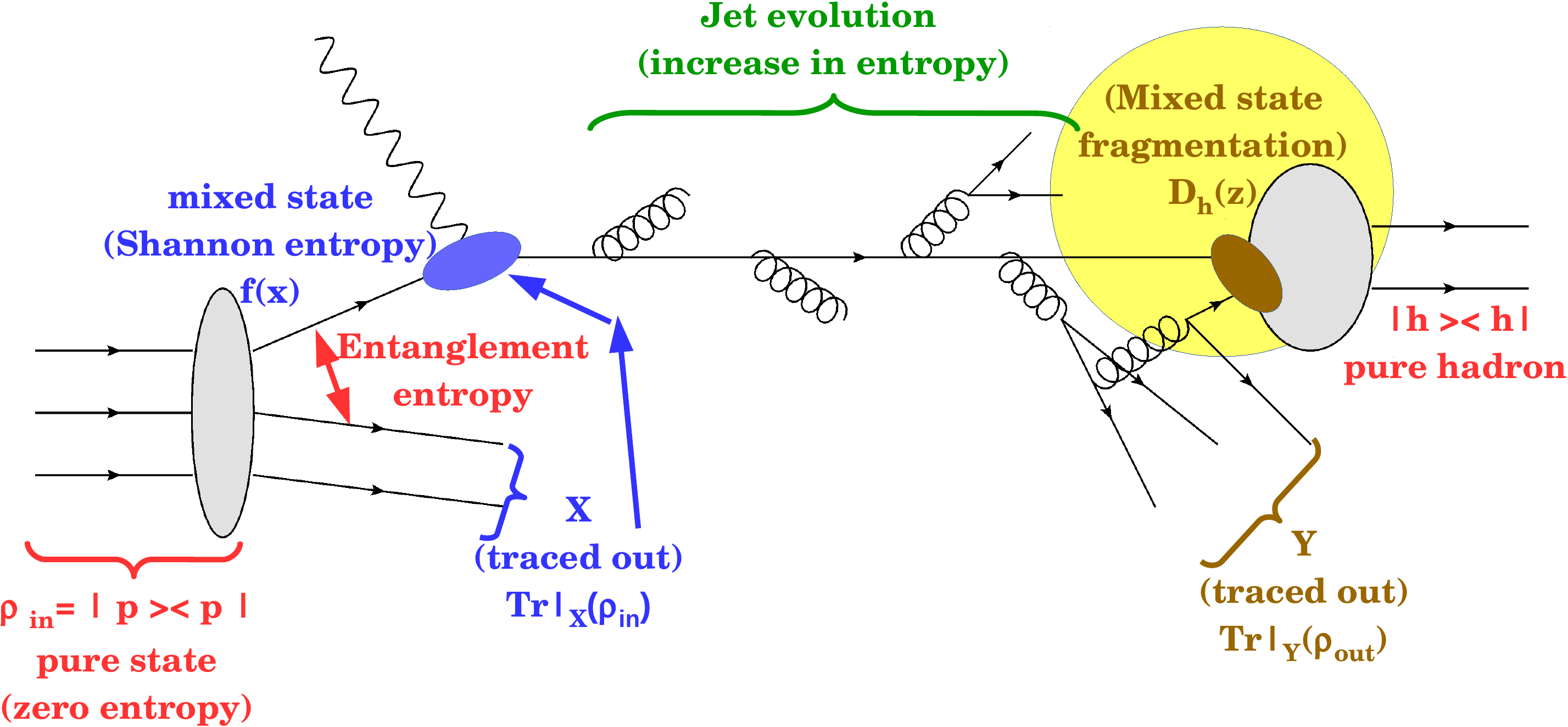}
	  \caption{
Illustration of the parts of a high-energy process for which entropy has been worked out.
}\label{globalscheme}
\end{figure}

Information theory is making inroads into high-energy physics: particularly, the flow of entropy in 
high-energy collisions and the decays produced therein has been examined at several stages. Figure~\ref{globalscheme}
schematizes various parts of a typical high-energy process that have been examined in the literature.
From left to right in the figure, the transformation of quantum entanglement entropy~\cite{Ehlers:2022oal} in the proton into classical Shannon entropy of the parton distribution function for the parton ejected in Deeply Inelastic Scattering (DIS) upon summing (tracing out) the proton remainder has been examined by Kharzeev and Levin~\cite{Kharzeev:2021yyf,Hentschinski:2022rsa}. 
The evolution of entropy in a jet is the purpose of  Neil and Waalewijn~\cite{Neill:2018uqw}. Here, one can think of the degrading (falling) parton $Q^2$ as analogous to a time advance in an irreversible process. 
Finally, to the right of the diagram, the parton fragments into an identified hadron and the Shannon entropy of the fragmentation function is the focus and first goal of this work as well as that of a companion article~\cite{Benito-Calvino:2022kqa} on
which it is based.  Other works cited therein and in~\cite{CarrascoMillan:2018ufj} have dealt with the entropy intrinsic to the distribution of the decaying products for the unstable hadrons formed in a collision, particularly of the Higgs boson, and other applications.

Let us quickly recall the concepts of Shannon entropy and Kullback-Leibler divergence. 
Given a system of $p$ elementary degrees of freedom  that take one of two possible values (bits),
the probability of a configuration in terms of the number  of bits of information yields a reasonable definition of the information as a logarithm of the probability:
$ p=  \left(\frac{1}{2} \right)^I \implies I := - \log_2(p)$.
Turning to the natural $e$-based logarithms and taking the expectation value of the information in the whole probability distribution, which for any function is $\langle f(p_i) \rangle = \sum_i p_i f(p_i)$, yields the Shannon entropy, 
\begin{equation} \label{ShannonS}
S= - \sum_i p_i \log p_i\ ,
\end{equation}
whose quantum version is of course the Von Neumann entropy  $S=-{\rm Tr} (\rho \log\rho) $.

Unfortunately, the entropy for a continuous probability distribution taken as the limit of a
simple discretization 
$ p(x),\ x\in[0,1] \to \left\{ p_1,p_2\dots p_i \dots p_N \right\}$; \ \ \ \ \ $\sum_i p_i =1$ 
fails: the computed entropy diverges upon increasing the number of points in the discretization, 
$S=-\sum_{i=1}^N \left (p_i\log p_i\right)  \xrightarrow{\color{blue} N\to \infty} \log N \to \infty  $. 
Therefore it is customary to adopt a generalization of Eq.~(\ref{ShannonS}) to a continuum distribution (that can now be a negative function),
\begin{equation} \label{Scontinuum}
S(F)\coloneqq - \int f(x)\ln f(x)\ud x \ .
\end{equation}

To make sense of this ``entropy''  we note that Eq.~(\ref{ShannonS}) is a particular case of the Kullback-Leibler divergence
quantifying how different the distribution $P$ is from a reference  $Q$,
\begin{equation}
 D_\mathrm{KL}(P\Vert Q)\coloneqq\sum_xp(x)\log\dfrac{p(x)}{q(x)}\ .
\end{equation}
and which admits a continuum generalization
\begin{equation} \label{KLcontinuum}
D_\mathrm{KL}(F\Vert G)=\int f(x)\log\dfrac{f(x)}{g(x)}\ud x\ .
\end{equation}

It is then obvious that Eq.~(\ref{Scontinuum}) is a particular case
$ S(F)\coloneqq - \int f(x)\ln f(x)\ud x $ in which  the KL divergence is taken 
to the uniform distribution (that with least information).

\begin{figure}[b!]
\begin{minipage}{0.66\textwidth}
\includegraphics[width=0.48\columnwidth]{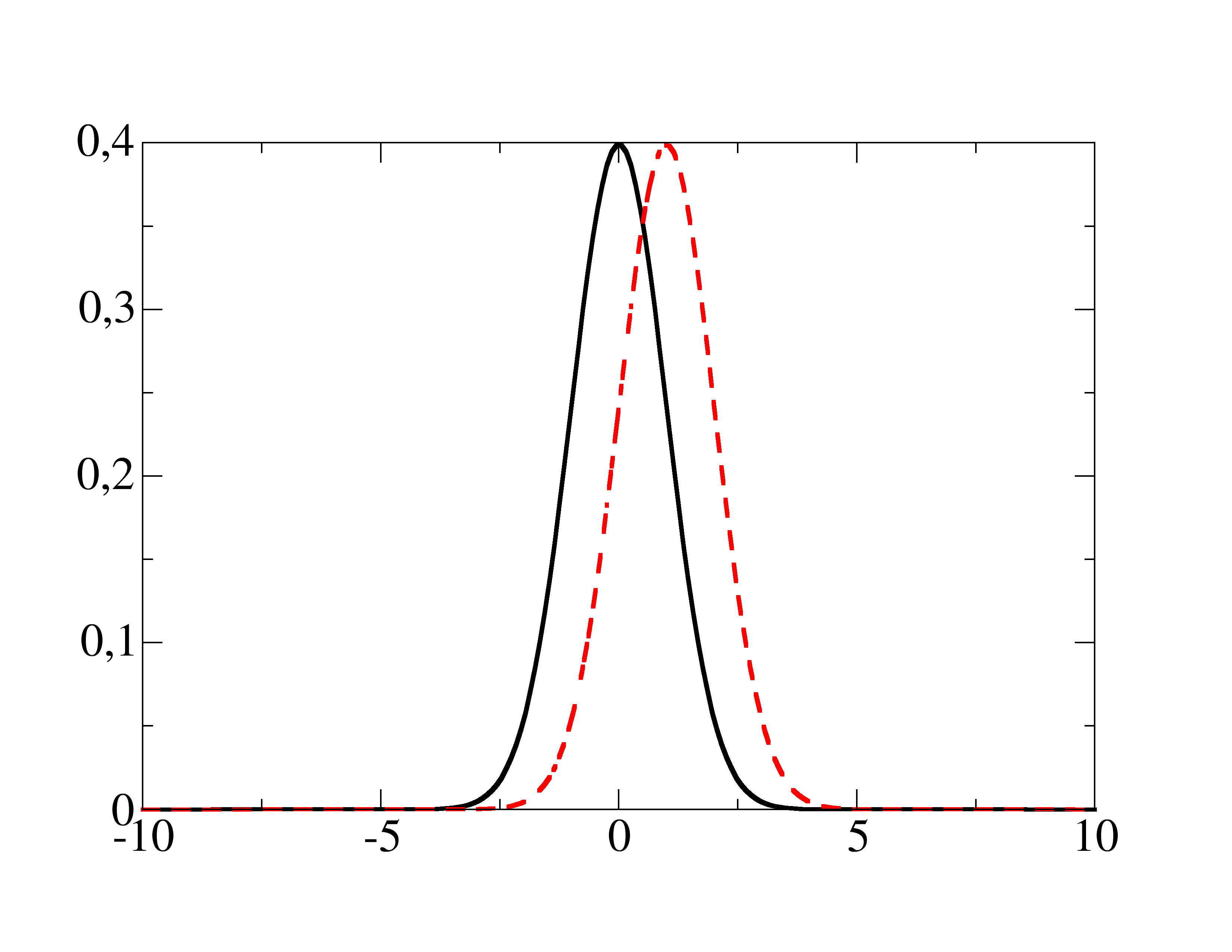}
\includegraphics[width=0.48\columnwidth]{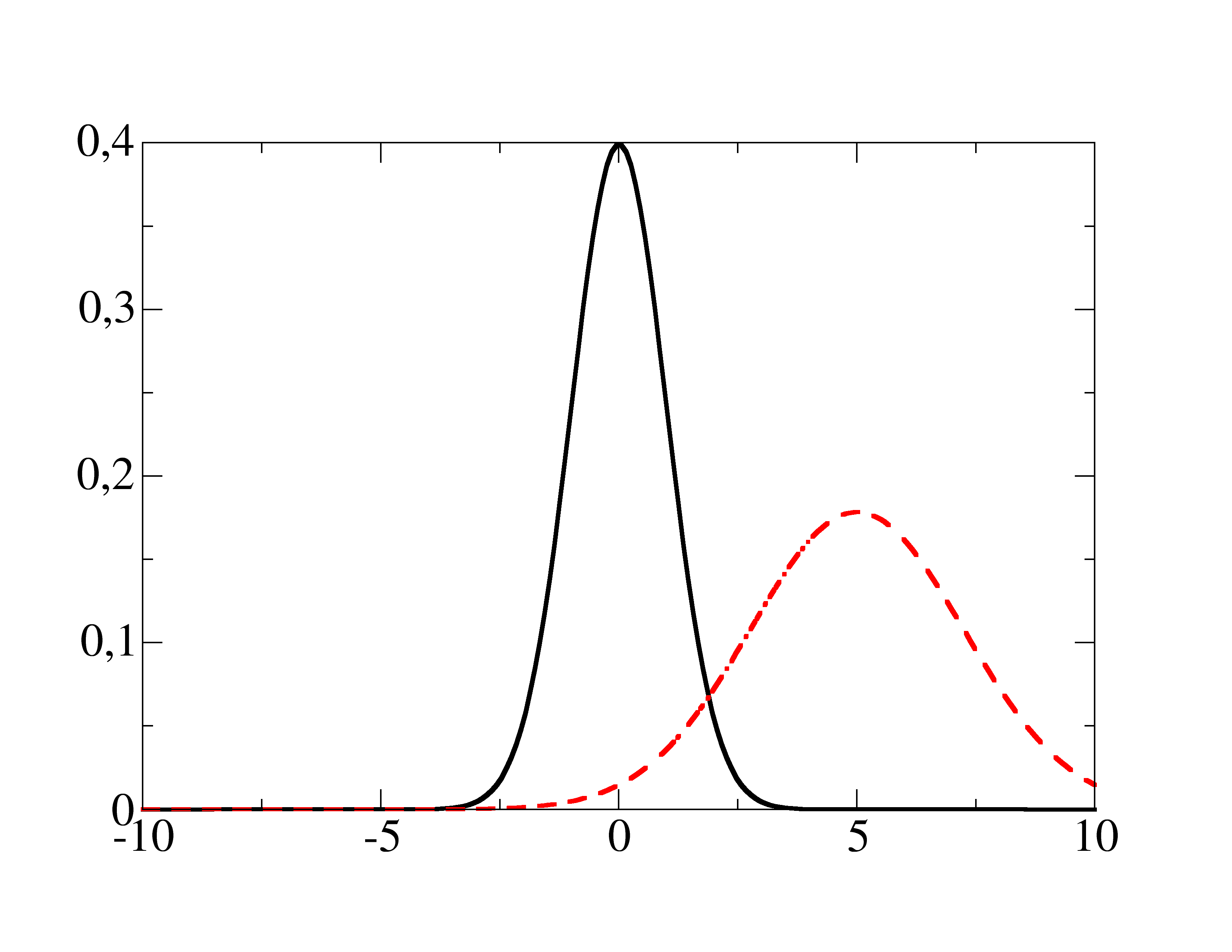}
\end{minipage}
\begin{minipage}{0.32\textwidth}
\caption{ Two pairs of Gaussian functions with widely different Kullback-Leibler divergences,
$D_{\rm KL} = 0.12 $ for the left pair of similar distributions, and $D_{\rm KL}= 1.63$ for the right one.
\label{fig:example}}
\end{minipage}
\end{figure}

To get a feeling for the size of the Kullback-Leibler divergence we plot two typical cases in figure~\ref{fig:example}.

\section{Crossing and relations between parton distribution and fragmentation functions}

Fragmentation functions and parton distribution functions play analogous roles in two processes related by crossing,
$e^-e^+\rightarrow hX$ and $e^- h \rightarrow e^-X$ respectively, where $h$ is an identified hadron and $X$ the remainder, unidentified, or at the parton level, $q\to hX$ and $h\to qX$. Table~\ref{tab:processes} exposes the analogy.
\begin{table}
\begin{minipage}{0.33\textwidth}\hfill
\caption{The analogy between fragmentation functions and parton distribution functions, showing the cross sections in terms of the structure functions at LO, the structure functions in terms of the fragmentation or distribution functions, and the respective momentum sum rules.
}\label{tab:processes}
\end{minipage}
\begin{minipage}{0.65\textwidth}
\begin{tabular}{|c|c|} \hline
Parton distributions & Fragmentation functions \\ \hline
& \\
 \small
$ \frac{d\sigma(e^- h \rightarrow e^-X)}{dxdy_{\rm l}}= 
\frac{2\pi \alpha^2}{Q^4}s[1+(1+y_{\rm l})^2 F_2(x)] $  &
 \small
$ \frac{d\sigma(e^-e^+\rightarrow hX)}{dz}= \frac{4\pi\alpha^2}{3Q^2}z^2 \hat{F_2}(z)$
 \\
& \\
 $F_2{(x)}=\sum_{i}e_i^2x[f_i(x)+\bar{f}_i(x)]$  &  $\hat{F_2}{(z)}=-3\sum_{q}e_q^2\frac{D^q(z)+D^{\bar{q}}(z)}{z^2}$ \\ 
& \\ 
 $\sum_{i}\int xf_i(x)dx=1$  &  $\sum_{h}\int_{0}^{1}zD_h^q(z)dz=1$ \\ \hline
\end{tabular}
\end{minipage}
\end{table}
The last row of the table reminds us of the generic momentum sum rule $\int dx p f(x) = P$ yielding the total momentum of the hadron (for the pdf) or the fragmenting parton (for the fragmentation function) that, after normalizing to 1 employing $p/P=x$, provides a convenient $xf(x)$ (or $zD(z)$ in the case of fragmentation) probability density.

As pointed out early on~\cite{Drell:1969jm},  DIS and fragmentation in $e^-e^+$ are related by analytic continuation due to crossing symmetry, which led Drell, Levy and Yan to propose the existence of a relation between pdfs and FFs, namely
\begin{equation}
D(z) = z\ f\left( \frac{1}{z} \right)\ .
\end{equation}
Such relation can run into problems at NNLO or higher in QCD, but it is certainly an asymptotic relation at the parton model limit. A more practical problem is that, because $z\in [0,1]$ is the momentum fraction of the fragmenting quark that the fragmented parton carried out, the argument of the parton distribution function, $\frac{1}{z}\geq 1$ is unphysical.

A second relation between fragmentation functions and pdfs is credited to Gribov and Lipatov, the ``reciprocity'' relation $3 D(z) = f(z)$, but its theoretical status seems to be rather that of a model~\cite{Barone:2000tx}. Barone, Drago and Ma revisited the issue and proposed, for  $z > 0.6$ the relation
\begin{equation}\label{BDG}
D(z)\simeq zf\left( 2-\frac{1}{z}\right)
\end{equation}
that is approximate save as $z\to 1$; but both functions have a physical argument and therefore the relation can be tested.
This is the second goal of the present article.

\section{Entropy as an indicator of progress in FF knowledge}

\begin{figure}
\includegraphics[width=\columnwidth]{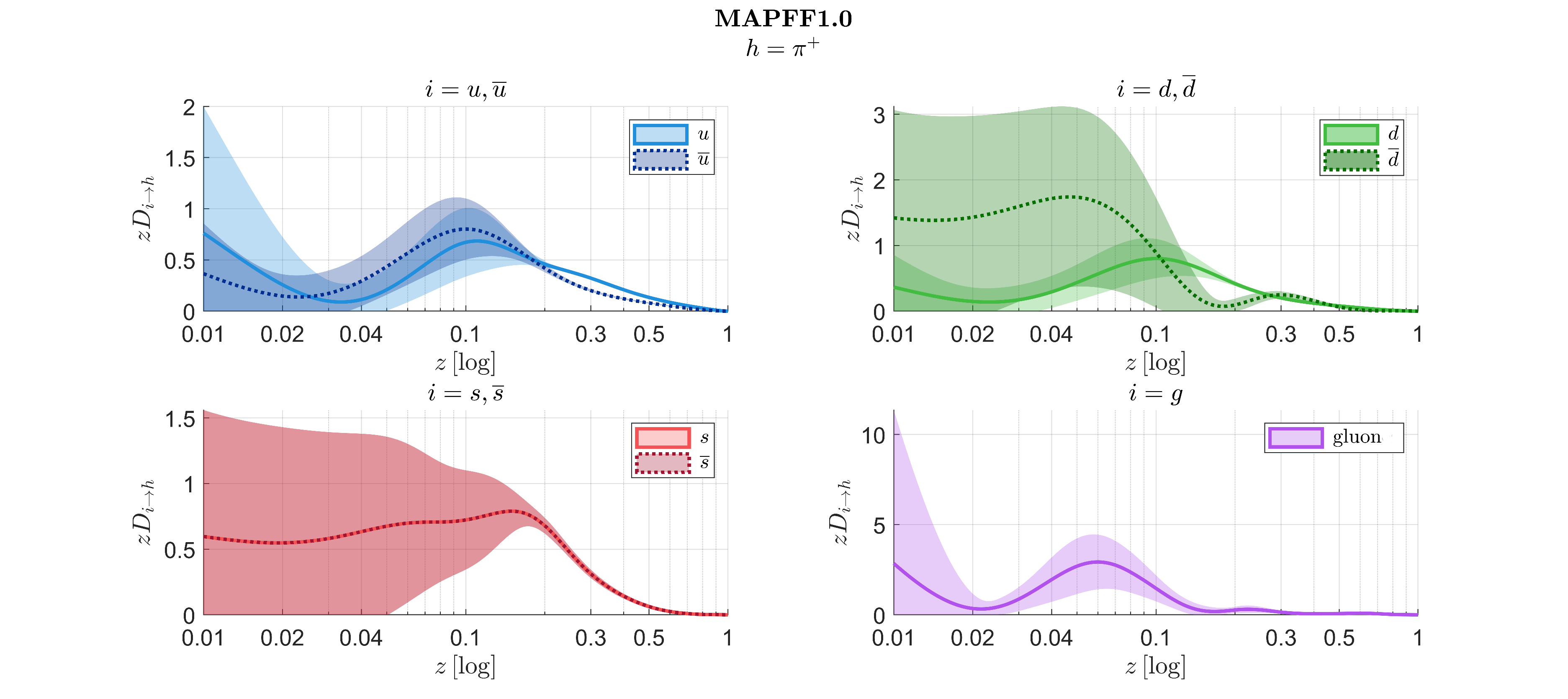}
\caption{\label{fig:FFpion} Fragmentation function of various partons to an identified pion, with uncertainty bands.}
\end{figure}

We employ fits to fragmentation functions based on neural networks such as those from the NNPDF or MAPFF
collaborations~\cite{Bertone:2017tyb,Khalek:2021gxf,AbdulKhalek:2022laj}. Figure ~\ref{fig:FFpion} exemplifies the typical sets with the fragmentation of different partons to the same hadron, the charged pion.

\begin{figure}
\includegraphics[width=\textwidth]{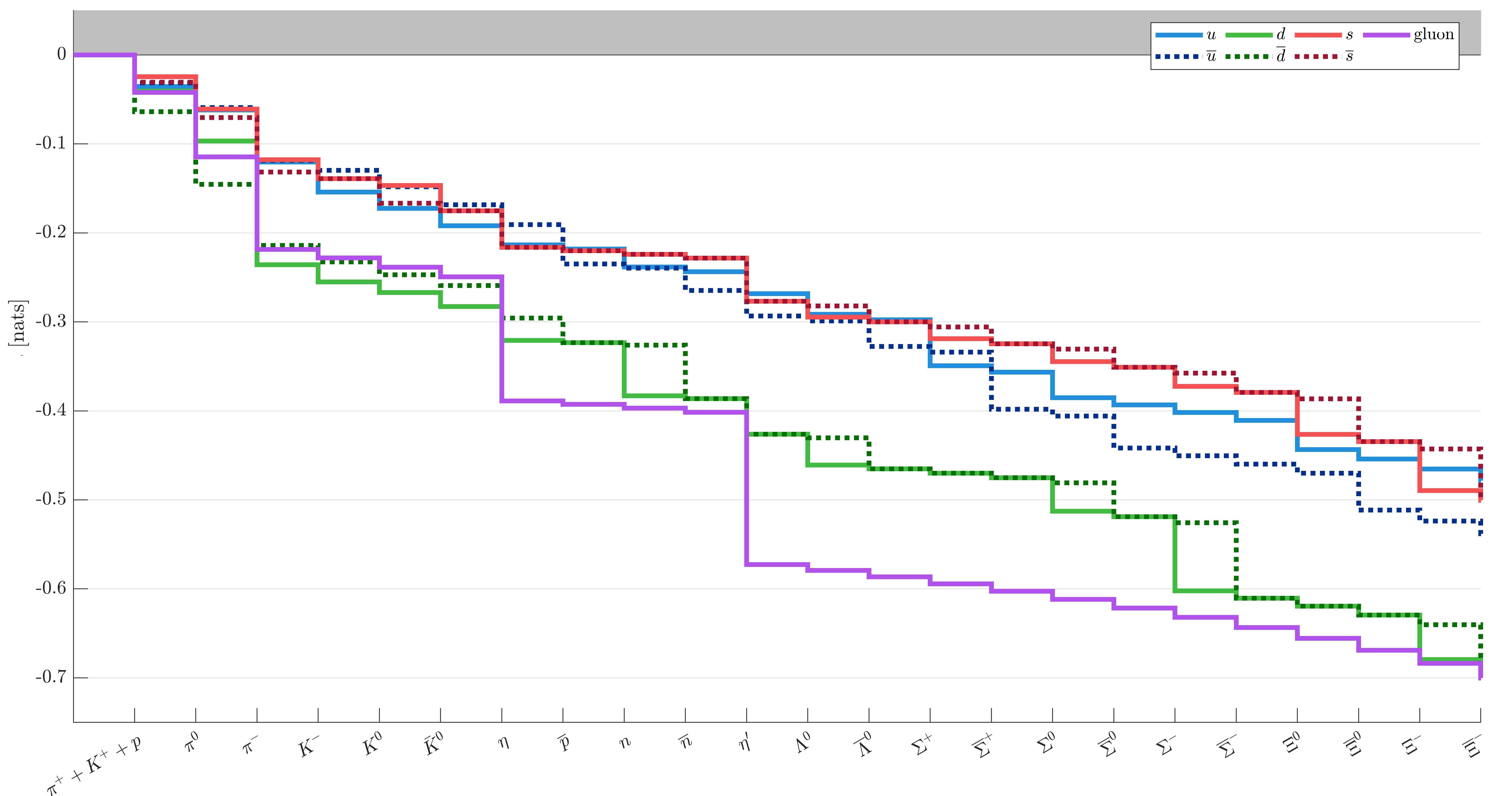}
\caption{Simulated improvement on the upper bound on $S$ as more fragmented hadrons are identified.}
\label{fig:boundonS}
\end{figure}

However, these (or any sets) do not saturate the sum over hadrons $h$ in the last row of table~\ref{tab:processes} necessary for the normalization: in practice,  $\pi^\pm$, $K^\pm$ and $p$ are provided, and little more.
What one can do with insufficient data is to renounce computing the entropy $S$ itself, and trying instead to obtain improvable upper bounds, by splitting 
$D_q(z) = D_q^{\rm measured}(z) + D_q^{\rm unknown}(z)$ satisfying the normalization
\begin{equation}
\underbrace{\sum_{h}}_{\rm measured}\int_{0}^{1}zD_h^q(z)dz=p<1\ \ \ \ \ \ \ \ 
\underbrace{\sum_{h}}_{\rm  unknown} zD_h^q(z) :=  1-p\ ;
\end{equation}
to maximize the entropy of the remainder, so that an upper bound to $S$ follows, we adopt a uniform distribution.
Then we can mimic what future advances in these FF data sets could bring about by adding an increasing number of hadrons,
as shown in figure~\ref{fig:boundonS}.

To prepare the simulation shown in the figure we have used standard quark model $SU(3)$ relations as well as a  valence/sea separation. For example, $D_{d\to K^0} $ (unknown) is set to $D_{u\to K^+}$  (that is already known in the mentioned FF sets). Also, the mixing angle $\theta(\eta,\eta')= -15.5^{\rm o}$ is taken from~\cite{Bramon:1997va}. A fully detailed table with all $q/h$ channels is given in~\cite{Benito-Calvino:2022kqa}.

It appears to us that this bound on Shannon's entropy $S$ is one number that quantifies the progress of any given set of Fragmentation Functions.

\section{Test of the Barone-Drago-Ma relation among $D$s and pdfs}

\begin{figure}
        \includegraphics[width=0.45\columnwidth]{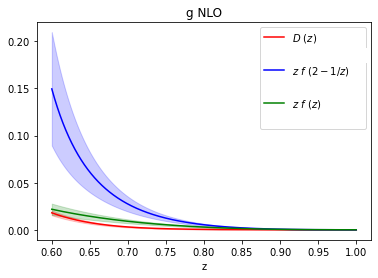} \ \
        \includegraphics[width=0.45\columnwidth]{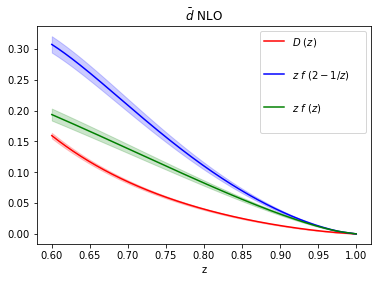}\\
        \includegraphics[width=0.45\columnwidth]{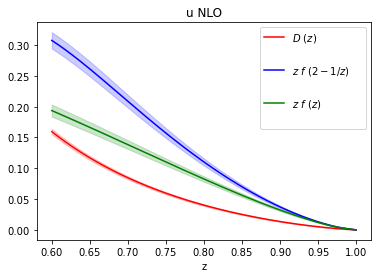} \ \
        \includegraphics[width=0.45\columnwidth]{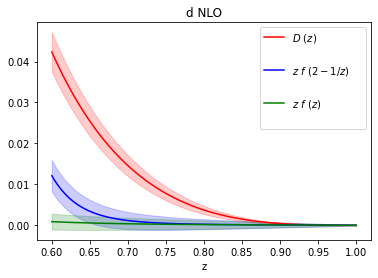}
\caption{\label{fig:JAM} Fragmentation functions and parton distribution function from the same set~\cite{Moffat:2021dji}
used to test the Barone-Drago-Ma relation.}
\end{figure}

Given the still large uncertainty band of the sets in the last section, here we employ more 
traditional template fits to $\pi^+$ data, taken at the scale $Q^2=100 \rm{GeV^2}$
and shown in figure~\ref{fig:JAM}. These are from the Jefferson Lab Angular Momentum Collaboration~\cite{Moffat:2021dji}.
(Using instead the proton's NNPDF parametrizations, $Q^2=100 \rm{GeV^2}$ we find similar results.)

One cannot tell from that figure alone whether similarities are due to the Barone-Drago-Ma relation being well satisfied
or simply due to the vanishing of all three functions plotted when $z\to 1$. 
Therefore, to discern the status of the relation, we quantify the separation of the functions via the Kullback-Leibler divergence, as shown in table~\ref{tab:KL}.
\begin{table} \centering
\caption{Values of the Kullback-Leibler divergence between the parton distribution function with argument $2-1/z$ as in Eq.~(\ref{BDG}), or with the further simplified $2-1/z\to z$, to the fragmentation function. (We have employed the NLO parametrizations of JAM coll. for $\pi^+$ fragmentation at scale $Q^2=100 \rm{GeV^2}$)\label{tab:KL}}
\begin{tabular}{||c c c||} 
    \hline
    Parton & $D_{KL}(zf_\pi(z)||D^\pi(z))$ & $D_{KL}(zf_\pi(2-1/z)||D^\pi(z))$      \\ [0.5ex] 
    \hline\hline
    gluon & 2,28 & 7,70 \\ 
    \hline
    up & 1,19  &  2,36\\
    \hline
    anti-down & 1,19 & 2,36  \\
    \hline
    down & 6,64 &  5,28  \\
    \hline
    \end{tabular}
\end{table}

The numbers in the table are obtained  from integrating over $z\in(0.7,1)$ where the Barone-Drago-Ma relation should be valid. We see that the divergences obtained are rather sizeable, so that the distributions are not very similar.

\begin{figure}\centering
\includegraphics[width=0.45\columnwidth]{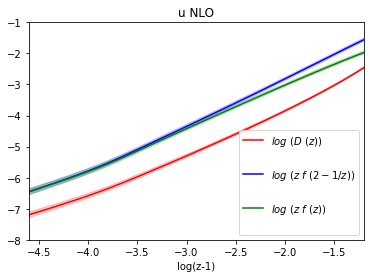}\ \
\includegraphics[width=0.45\columnwidth]{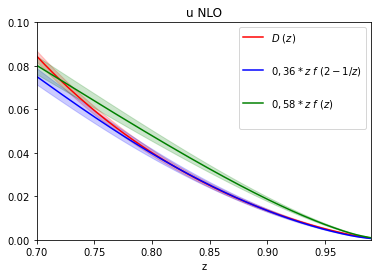}
\caption{\label{fig:straightlines}
The functions entering the Barone-Drago-Ma relation in a logarithmic plot showing near parallelism (left) and the nearly equal rescaled functions in a linear plot (right).
}
\end{figure}

However,  the JAM fits do seem to favor some relation between FFs and pdfs; we exemplify with $f_\pi^u(x)$ and $D_u^\pi(z)$, 
in a log-log plot in figure~\ref{fig:straightlines}.
The figure clearly shows near parallelism, meaning that the Barone-Drago-Ma relation is off by a multiplicative constant only!
In fact, the further approximating $f(2-1/z)\to f(z)$ is not so bad either. Rescaling the pdfs by an arbitrary constant puts them on top of the fragmentation function $D(z)$ as seen in the right plot.
This means that the three functions have very similar power-law exponent upon coming into the $z\to 1$ endpoint (this is reminiscent of the Drell-Yan-West relation between the power-laws of the elastic form factors and the parton distribution functions), 
with respective values 1.48 for $D(z)$, 1.52 for $zf(2-\frac{1}{z})$ and 1.37 for $zf(z)$.

\section{Conclusions and outlook}

In conclusion, we have newly deployed information entropy and the Kullback-Leibler divergence for
studying  fragmentation functions in high-energy collisions. 

Our two findings are that the entropy $S$ is one simple number that can quantify progress on $FF$ knowledge;
and that while the  relations $pdfs \leftrightarrow FFs$ are not well satisfied with current sets,
suggesting perhaps that the Barone-Drago-Ma relation is trivial and good only at $z=1$ exactly,
more investigation is warranted as the JAM collaboration fits seem to suggest that they have similar slopes near the endpoint.

 If we were to take a guess at what the main experimental difficulty might be, other than the usual technical problems
in identifying hadrons in a collider setting, we would observe that the parton distribution function is certainly measured on the said hadron, be it the  proton or the charged pion, by target preparation in a DIS experiment. On the contrary, the fragmentation function must receive contribution from excited baryons that decay to the proton before reaching any detector, $N^*,\Delta^*\to p\dots$ typically within the few femtometers characteristic of the strong force. These resonances thus feed down to the fragmentation function $D(z)$, but not to the parton distribution function.

\section*{Acknowledgments}
This project has received funding from the European Union's Horizon 2020 research and innovation programme under grant agreement No 824093; grants  MICINN: PID2019-106080GB-C21, PID2019-108655GB-I00/AEI/10.13039/501100011033    (Spain); UCM research group 910309 and the IPARCOS institute.




%
\bibliography{refsLlanes}

\begin{thebibliography}{13}

\bibitem{Ehlers:2022oal}
P.J. Ehlers (2022), \texttt{2209.09867}

\bibitem{Kharzeev:2021yyf}
D.E. Kharzeev, E.~Levin, Phys. Rev. D \textbf{104}, L031503 (2021),
  \texttt{2102.09773}

\bibitem{Hentschinski:2022rsa}
M.~Hentschinski, K.~Kutak, R.~Straka (2022), \texttt{2207.09430}

\bibitem{Neill:2018uqw}
D.~Neill, W.J. Waalewijn, Phys. Rev. Lett. \textbf{123}, 142001 (2019),
  \texttt{1811.01021}

\bibitem{Benito-Calvino:2022kqa}
G.~Benito-Calvi\~no, J.~Garc\'\i{}a-Olivares, F.J. Llanes-Estrada (2022),
  \texttt{2209.13225}

\bibitem{CarrascoMillan:2018ufj}
P.~Carrasco~Mill\'an, M.A. Garc\'\i{}a-Ferrero, F.J. Llanes-Estrada,
  A.~Porras~Riojano, E.M. S\'anchez~Garc\'\i{}a, Nucl. Phys. B \textbf{930},
  583 (2018), \texttt{1802.05487}

\bibitem{Drell:1969jm}
S.D. Drell, D.J. Levy, T.M. Yan, Phys. Rev. \textbf{187}, 2159 (1969)

\bibitem{Barone:2000tx}
V.~Barone, A.~Drago, B.Q. Ma, Phys. Rev. C \textbf{62}, 062201 (2000),
  \texttt{hep-ph/0011334}

\bibitem{Bertone:2017tyb}
V.~Bertone, S.~Carrazza, N.P. Hartland, E.R. Nocera, J.~Rojo (NNPDF), Eur.
  Phys. J. C \textbf{77}, 516 (2017), \texttt{1706.07049}

\bibitem{Khalek:2021gxf}
R.A. Khalek, V.~Bertone, E.R. Nocera, Phys. Rev. D \textbf{104}, 034007 (2021),
  \texttt{2105.08725}

\bibitem{AbdulKhalek:2022laj}
R.~Abdul~Khalek, V.~Bertone, A.~Khoudli, E.R. Nocera, Phys. Lett. B
  \textbf{834}, 137456 (2022), \texttt{2204.10331}

\bibitem{Bramon:1997va}
A.~Bramon, R.~Escribano, M.D. Scadron, Eur. Phys. J. C \textbf{7}, 271 (1999),
  \texttt{hep-ph/9711229}

\bibitem{Moffat:2021dji}
E.~Moffat, W.~Melnitchouk, T.C. Rogers, N.~Sato (Jefferson Lab Angular Momentum
  (JAM)), Phys. Rev. D \textbf{104}, 016015 (2021), \texttt{2101.04664}

\end{thebibliography}

\end{document}